\documentclass[doublespace,onecolumn]{SSA-SRL}

\usepackage{amsmath,amsfonts}
\usepackage{algorithm}
\usepackage{changepage}
\usepackage{hyperref}

\newcommand{\argmin}{\mathop{\rm arg~min}\limits}

\citethisauthor{Agata~R., S.~Baba, A.~Nakanishi and Y.~Nakamura}
\doi{00.0000/000000000}
\recdate{00 Month 0000}

\begin{document}

\title{HypoNet Nankai: Rapid hypocenter determination tool for the Nankai Trough subduction zone using physics-informed neural networks}

\author[*1\orc{0000-0002-4866-8267}]{Ryoichiro Agata}
\author[1\orc{0000-0002-5828-8030}]{Satoru Baba}
\author[1\orc{0000-0001-5097-7248}]{Ayako Nakanishi}
\author[1\orc{0000-0001-6750-2791}]{Yasuyuki Nakamura}

\affil[1]{Japan Agency for Marine-Earth Science and Technology (JAMSTEC), Yokohama, Kanagawa, Japan}{\auorc[https:// orcid.org/]{0000-0002-4866-8267}{(FA)}\auorc{0000-0002-5828-8030}{(SA)}}
\corau{*Corresponding author: agatar@jamstec.go.jp}

\maketitle
\begin{adjustwidth}{2cm}{}
\section{Abstract}
\textbf{
Accurate hypocenter determination in the Nankai Trough subduction zone is essential for hazard assessment and advancing our understanding of seismic activity in the region. 
Therefore, a handy hypocenter determination tool incorporating a realistic three-dimensional (3D) velocity structure, accessible to the scientific community, is beneficial. 
In this study, we developed HypoNet Nankai, a rapid hypocenter determination tool based on a physics-informed neural network (PINN) emulator (surrogate model) for travel time calculations. 
This tool leverages a PINN trained to predict P-wave travel times between arbitrary underground sources and surface receivers with a realistic 3D P-wave velocity structure model of the Nankai Trough subduction zone that incorporates marine seismic survey data. 
The PINN embeds physical laws, namely, the Eikonal equation, directly into the loss function of training and circumvents the need for labeled training data. 
To address the training challenges posed by small-scale features in the velocity model, we employed a simple domain decomposition approach and Fourier feature embedding. 
Once trained, the PINN immediately infers the P-wave travel time, enabling rapid hypocenter determination. 
The data size required to store NNs for travel time calculations is significantly  smaller than those of conventional travel-time tables. 
HypoNet Nankai provides high flexibility for addition of new observation points. 
We verified HypoNet Nankai by comparing its performance with a widely used grid-based numerical method for forward travel time calculations and synthetic hypocenter determination. 
In both tests, HypoNet Nankai provided results consistent with those for the conventional method.  
HypoNet Nankai offers a rapid, accurate, and easy-to-use hypocenter determination method for the Nankai Trough subduction zone, with greater data efficiency and extendibility compared to conventional approaches.
The tool is publicly accessible at \url{http:} (under preparation). 
}
\end{adjustwidth}


\section{Introduction}

The Nankai Trough subduction zone is one of the most active subduction zones in the world, having experienced repeated megathrust earthquakes with magnitudes ranging from M8-9 \cite[]{Ando1975}. 
The 30-year occurrence probability of such a large Nankai megathrust earthquake is estimated to be 70–80\% \cite[]{ERC2024}.
Therefore, achieving accurate hypocenter determination in such an active subduction zone to precisely understand the seismic activity is crucial for both hazard assessment and advancing scientific understanding.

The foundation of hypocenter determination analysis lies in the theoretical calculations of travel times from the seismic source in the Earth's interior to seismological observation points on the Earth's surface. 
Classic methods based on simple velocity structure models that are easy to handle (e.g., one-dimensional (1D) structures), have been widely used \citep[for example][]{Klein1978USGS,Hirata1987}. 
However, it is evident that incorporating more realistic three-dimensional (3D) velocity structure models has a substantial impact on the accuracy of hypocenter determination, particularly in regions with complex underground seismic velocity structures such as subduction zones and volcanoes \citep[for example,][]{Lomax2001GJI,Johnson2002BSSA,Nakano2015,Katsumata2015BSSA}. 
Therefore, it is crucial to prepare appropriate 3D velocity structure models and perform 3D travel time calculations aiming at accurate hypocenter determination.

Although establishing an appropriate 3D velocity structure model is challenging, research on velocity structures has made significant progress in the Nankai Trough region.
 Consequently, multiple realistic 3D velocity structure models have been proposed for this region \cite[]{Koketsu2009,Koketsu2012,Yamamoto2017EPSL,Nakanishi2018,Arnulf2022NatGeo}. 
3D travel time calculations have been established using methods such as ray tracing \cite[]{Julian1977JG,Um1987BSSA}, grid-based shortest path algorithms \cite[]{Moser1991Geophysics}, and grid-based finite-difference calculations \cite[]{Sethian1996,Zhao2005}, with many open-source programs available for these approaches \cite[for example,][]{White2020SRL,Giroux2021}. 
Nevertheless, it is still common to observe simple structures, such as 1D models, being used in many studies, suggesting significant 
computational costs and effort required for 3D travel time calculations in practice. 
Making accessible tools with which users can easily obtain results is essential for facilitating the adoption and widespread use of 3D velocity structure models for travel-time calculations and hypocenter determination.

A traditional approach that can be considered a solution is to precompute and store travel-time tables from underground grid points, which are assumed to be hypothetical hypocenter locations at each observation station \cite[for example,][]{Katsumata2015BSSA}. 
By utilizing these tables, users can interpolate and calculate travel times from the desired hypocenter locations. 
However, this approach has several limitations. For instance, dense grid points are required to ensure accurate interpolation, which results in an enormous amount of data.
In addition, when new observation stations are added, especially in high-density settings, such as distributed acoustic sensing (DAS), additional costly travel time calculations are required. 
Therefore, this approach can be considered suboptimal for tools available to the public.
A more modern approach involves machine learning models, such as deep neural networks (DNN), to produce an emulator (surrogate model) that bypass expensive calculation in modeling the nonlinear relationship between inputs (i.e., source and receiver locations) and outputs (i.e., travel times).
A trained DNN-based emulator enables travel time to be obtained rapidly via inference in forward pass.
The amount of data required to store the DNN is much smaller than that required for a travel-time table. 
However, when utilizing such machine learning methods, it is necessary to prepare a large amount of labeled  training data consisting of input and output pairs. 
This eventually requires an enormous number of theoretical 3D travel-time calculations. 

Recently, physics-informed neural networks (PINN) \cite[]{Raissi2019} have gained attention as a new approach to solving partial differential equations (PDEs). 
PINN can incorporate physical laws described by PDEs, such as the Eikonal equation for travel time calculation, into a loss function for training DNNs, eliminating the need for labeled training data. 
This allows the development of a neural network model that captures the nonlinear physical relationships between the source location, receiver location, and travel time without requiring the preparation of labeled data.
PINN for the rapid calculation of travel time has been successfully developed \cite[]{Smith2021,Waheed2021PINNeik,Grubas2023} and applied to developing emulators of travel time calculations for a real-world velocity structure at a global scale \cite[GlobeNN][]{Taufik2023SciRep} and in the Southern California region \cite[]{Smith2022}. 
The latter was further integrated into a Bayesian hypocenter determination tool called HypoSVI. 
Although these NN-based models have been verified based on the residuals of the Eikonal equation, their performance in terms of accuracy of inferred travel times and hypocenter determination has not been assessed.

To address the need for rapid and accurate hypocenter determination in the Nankai Trough subduction zone,  we developed HypoNet Nankai, a rapid hypocenter determination tool that employs physics-informed neural networks. This tool utilizes a PINN-based emulator model trained to predict travel times between arbitrary source and receiver pairs within a 3D P-wave velocity structure model of the Nankai Trough subduction zone \cite[]{Nakanishi2018}. 
This velocity model is based on previous marine active-source seismic data and provides the foundation for the tool’s application. 
We tested the accuracy of the PINN-based emulator and HypoNet Nankai by comparing them with a grid-based numerical calculation method for forward travel time calculations and synthetic hypocenter determination problems, respectively.

\section{Theory of PINN for travel time prediction}

We first explain the PINN formulation for travel time calculation \cite[]{Smith2021,Waheed2021PINNeik,Grubas2023}, which was slightly modified by \cite{Agata2023}. 
The Eikonal equation relates the spatial derivative of the travel time field to the velocity structure as follows:
\begin{align}
|\nabla T(\mathbf{x},\mathbf{x}_{s})|^{2} &= \displaystyle \frac{1}{v^{2}(\mathbf{x})}, \quad \forall\,\mathbf{x} \in \Omega \label{eqn:eikonal}\\
T(\mathbf{x}_{s}, \mathbf{x}_{s}) &= 0,\label{eqn:PC}
\end{align}
where $\Omega$ is an $\mathbb{R}^d$ domain, $d$ is the space dimension, $T(\mathbf{x},\mathbf{x}_{s})$ is the travel time at the point $\mathbf{x}$ from the source $\mathbf{x}_s$, $v(\mathbf{x})$ is the velocity defined in $\Omega$, and $\nabla$ denotes the gradient operator with respect to $x$. 
The second equation defines the point-source condition. 
To avoid singularities in this condition, previous studies modeling travel time using PINN have introduced the following factored form \cite[]{Smith2021,Waheed2021PINNeik}: 
\begin{equation}
T(\mathbf{x},\mathbf{x}_{s})=T_{0}(\mathbf{x},\mathbf{x}_{s}) \tau(\mathbf{x},\mathbf{x}_{s})
\end{equation}
where $T_{0}$ is defined as:
\begin{equation}
T_{0}(\mathbf{x},\mathbf{x}_{s})=\left|\mathbf{x}-\mathbf{x}_{s}\right|.
\end{equation}
This factorization automatically satisfies the point-source condition.
We introduce a NN to predict the travel time for a given velocity structure. 
The NN constructs a function $f_{T}$ characterized by the weight parameters $\boldsymbol{\theta}$. 
We define the NN-based function to approximate travel time as
\begin{align}
T(\mathbf{x},\mathbf{x}_{s}) 
&\simeq f_{T}(\mathbf{x},\mathbf{x}_{s},\boldsymbol{\theta})\\
&= T_{0}(\mathbf{x},\mathbf{x}_{s})/f_{\tau^{-1}}(\mathbf{x},\mathbf{x}_{s},\boldsymbol{\theta}),\label{eqn:T}
\end{align}
where $f_{\tau^{-1}}$ is the output of the NN used to approximate $1/\tau(\mathbf{x},\mathbf{x}_{s})$ instead of directly approximating $\tau$. 
We use fully connected feedforward networks with Fourier feature embedding \cite[]{Tancik2020,Hennigh2021} as explained in ``Strategy for accurate training of PINN'' to implement $f_{\tau^{-1}}$.
Additional operations are applied to normalize the input and output of NNs, improve the convergence performance, and set the upper and lower limits of the final output values.
Furthermore, the reciprocity condition (i.e., $T(\mathbf{x},\mathbf{x}_{s})=T(\mathbf{x}_{s},\mathbf{x})$) is imposed following \cite{Grubas2023} to improve the convergence of the solution to the Eikonal equation. 

The NNs are trained using the following loss functions:
\begin{equation}
L(\boldsymbol{\theta}) = \frac{1}{N_{\rm c}} \sum_{i=1}^{N_{\rm c}}\left(v(\mathbf{x}_{\rm c}^{(i)})-\frac{1}{|\nabla f_{T}(\mathbf{x}_{\rm c}^{(i)},\mathbf{x}_{\rm s}^{(i)}; \boldsymbol{\theta})|} \right)^{2}, \label{eqn:loss}
\end{equation}
where $N_c$ is the number of collocation- and source–point pairs.
$\mathbf{x}_{\rm c}$ and $\mathbf{x}_{\rm s}$ are the coordinates of the collocation and source points, respectively. 
The collocation points were set as the evaluation points for the PDE residuals \cite[]{Raissi2019}. 
The loss function, which incorporates physics-informed constraints, consists of the sum of the squared residuals for each pair of source and collocation points.  
$\mathbf{x}_{\rm c}^{(i)}$, which corresponds to $\mathbf{x}$ in Equation \ref{eqn:eikonal}, must cover the entire domain to ensure that the Eikonal equation is satisfied everywhere, regardless of how limited the distribution of receivers is to a small domain. 
In contrast, the distribution of $\mathbf{x}_{\rm s}^{(i)}$ in Equation \ref{eqn:loss}, which corresponds to $\mathbf{x}_s$ in Equation \ref{eqn:eikonal}, can be taken arbitrarily for the domain of interest regarding the source distribution.  
$f_{T}(\mathbf{x}_{\rm c},\mathbf{x}_{\rm s}; \boldsymbol{\theta}^{*})$ with the optimized NN weight parameters $\boldsymbol{\theta}^{*}= \argmin_{\boldsymbol{\theta}} L(\boldsymbol{\theta})$ can serve as an emulator, which can infer travel time between any in-distribution receiver-source pair. 

Our objective was to develop a PINN-based emulator that instantly infers the travel time between a source point anywhere in the given 3D velocity model and a receiver point anywhere on the Earth's surface within the velocity model domain. 
The distribution of the receiver points is considerably more limited than that of the source points. 
However, as stated previously, training $f_{T}$ is easier when the distribution of the source points is limited, which is the opposite of our case. 
Leveraging the reciprocity of the sources and receivers in the Eikonal equation, we used the Earth’s surface points as $\mathbf{x}_{\rm s}^{(i)}$ and the points from the velocity model as $\mathbf{x}_{\rm c}^{(i)}$ in the training stage (Fig. \ref{fig:VSMNN}). 

\section{Three-dimensional P-wave velocity structure model \& bathymetric data}

\begin{figure*}
\begin{center}
\begin{small}
\includegraphics[clip, width=16cm, bb = 0 0 779 432]{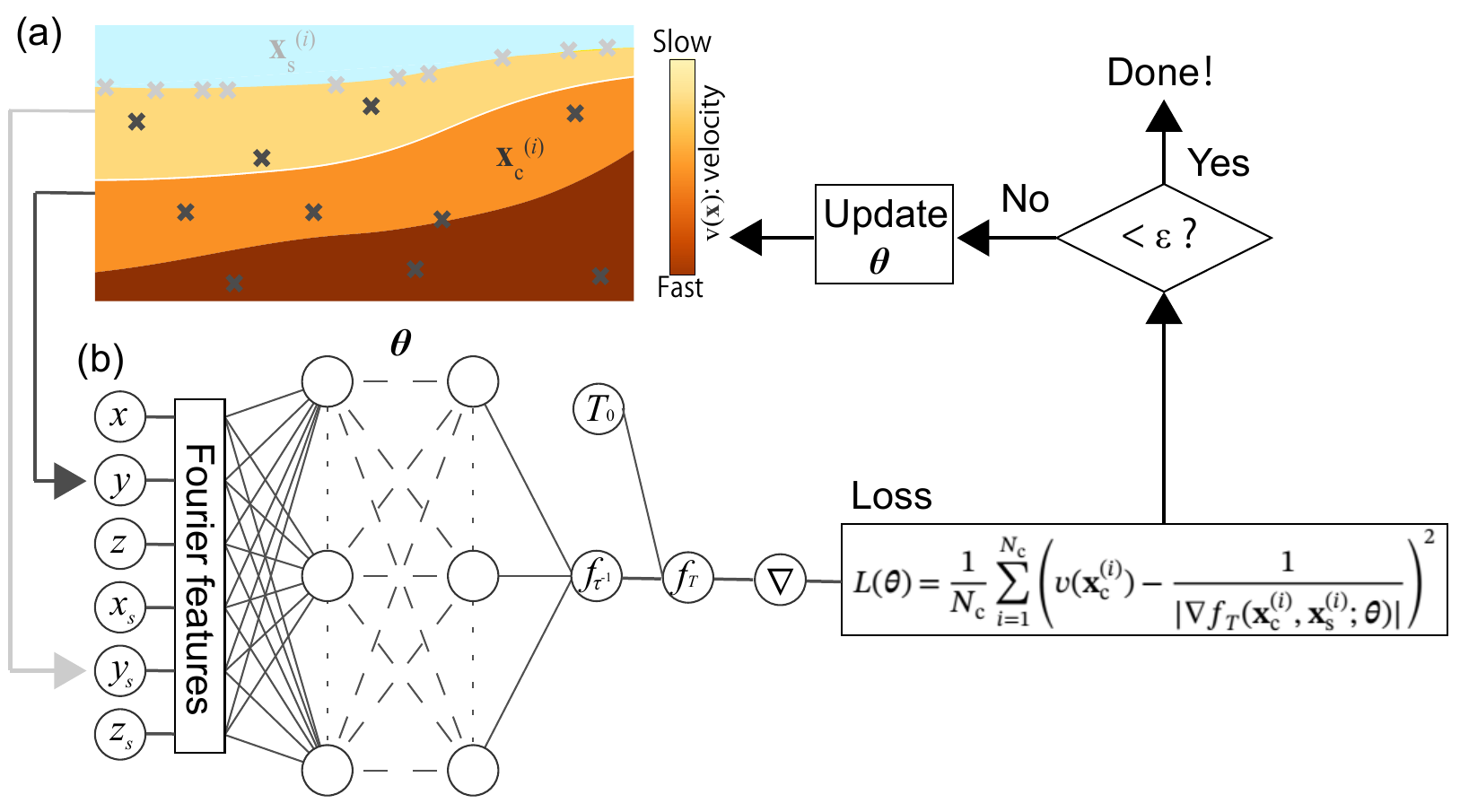}
\end{small}
\end{center}
\caption{Overview of training PINN-based emulator for travel time calculations using velocity structure model. (a) A 2D schematic of sampling of training points. Dark- and light-gray cross marks indicate the collocation and Earth's surface points ${\bf x}_c$ and ${\bf x}_s$, respectively. The light blue domain indicates seawater. 
(b) Schematic view of the neural network formulation and training based on the physics-informed loss function. Although $T_0$ is also a function of ${\bf x}=(x, y, z)^{\rm T}$ and ${\bf x_{\rm s}}=(x_{\rm s}, y_{\rm s}, z_{\rm s})^{\rm T}$, we do not describe this dependency here for simplicity. The dark- and light-gray points ${\bf x}_c$ and ${\bf x}_s$ in (a) are used as the input coordinates ${\bf x}$ and ${\bf x_{\rm s}}$ of the neural network, leveraging the reciprocity of the Eikonal equaion. }
\label{fig:VSMNN}
\end{figure*}

\begin{figure}
\begin{center}
\begin{small}
\includegraphics[clip, width=8cm, bb = 0 0 388 573]{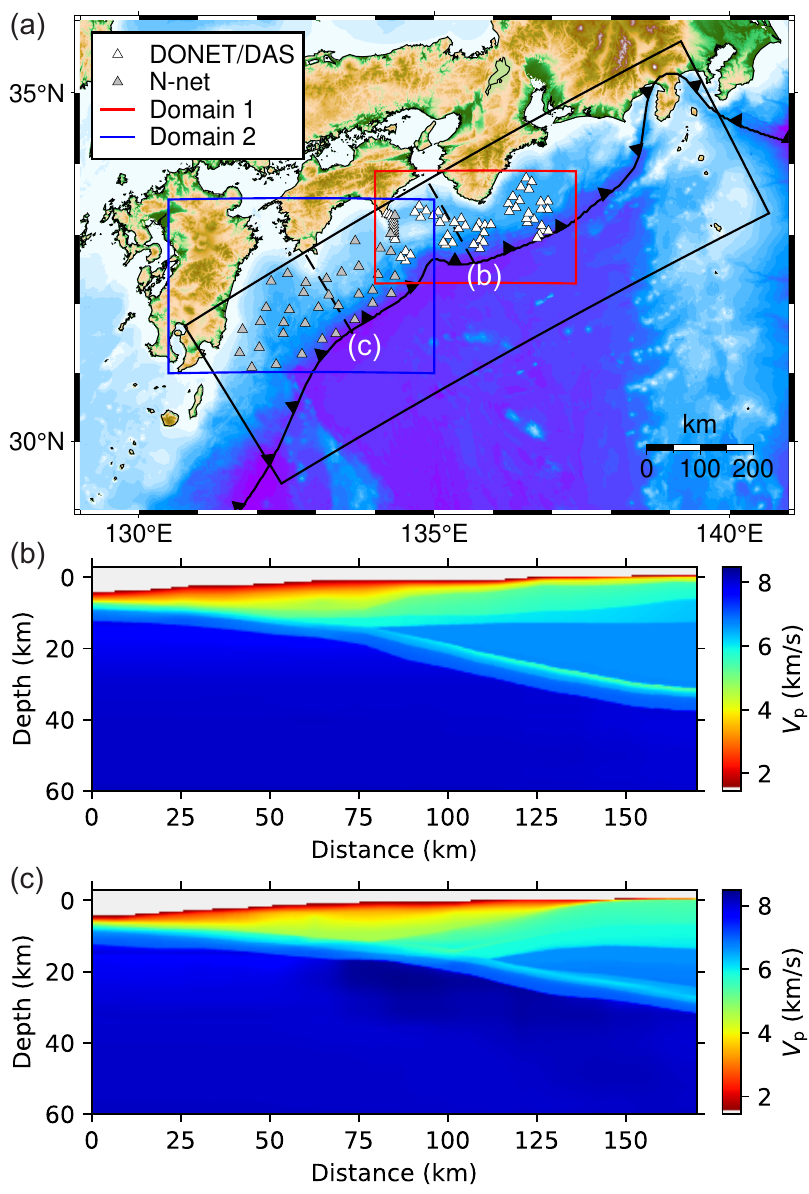}
\end{small}
\end{center}
\caption{(a) Map of the study area. The black solid rectangle denotes the modeling domain of the PINN employed in HypoNet Nankai. Domains 1 and 2 are the domains visualized in Fig. \ref{fig:hypo} (a) and (b), which show the results of the numerical experiments of hypocenter determination using the locations of the DONET/DAS and N-net observation points, respectively. 
(b) and (c) P-wave velocity profiles along dashed black lines plotted in (a). }
\label{fig:map}
\end{figure}

We used the P-wave velocity model, which covers the entire domain of the Nankai Trough subduction zone proposed by \cite{Nakanishi2018} (hereafter called N2018 model), to develop a NN model for travel time calculations. 
This model was created by merging two-dimensional (2D) P-wave velocity profiles obtained from  previous wide-angle seismic reflection surveys using OBS arrival time data \cite[]{JAMSTEC2004} and a 3D velocity model obtained by seismic tomography using natural earthquakes \cite[]{Yamamoto2013,Yamamoto2014}.
Because the N2018 model prioritizes the accurate modeling of offshore structures, this study focuses on developing a PINN specifically for travel-time calculations in the offshore region, which covers a horizontal area of 900\,km$\times$300\,km and a depth extent of 60\,km (Fig. \ref{fig:map}). 
We prepared $\mathbf{x}_{\rm c}^{(i)}$ based on 3D point cloud data comprising the latitude, longitude, depth, and P-wave velocity, which constitute the N2018 model (Fig. \ref{fig:VSMNN} (a)). 
The point locations are structured only in the vertical direction. 
We transformed them into a topocentric cartesian coordinate system, the origin of which was located at (136$^{\circ}$N, 33$^{\circ}$E, 0), without using map projection. 
The geoid data from EGM2008 \cite[]{Pavlis2012} were used for this transformation. 
Furthermore, we translated and rotated the system to facilitate the use of the calculation grid in the grid-based travel time calculations used in the comparisons. 
On average, the N2018 model sampled points every 0.5\,km in the horizontal direction and 0.1\,km in the vertical direction. 
Although dense sampling in the vertical direction helps capture sudden velocity jumps, such as those at the Moho, it oversamples other regions with moderate velocity changes. 
In the regions where the vertical velocity gradient was less than $0.1s^{-1}$, we downsampled the points to achieve a grid spacing of 1\,km.
This approach enables the optimization of computational resources by focusing on learning the travel time function in regions with significant vertical velocity changes. 
The total number of data points after this approach was approximately 150 million, all of which were used for the mini-batch training of the PINN. 
Because grid-based travel time calculation methods can provide reference solutions for verification in our target calculation, we did not divide the data for validation and testing, which is typical in machine learning training methods. 

We used GEBCO Gridded Bathymetry Data \cite[]{Gebco2023} to compose the data of the receiver locations that were introduced into the training as $\mathbf{x}_{\rm s}^{(i)}$ (Fig. \ref{fig:VSMNN} (a)). 
Similar to the velocity model, we transformed the original data consisting of latitude, longitude, and elevation into the cartesian coordinate system. 
When sampling the data, we randomly generated points in the horizontal domain and obtained the vertical positions of the points by nearest neighbor interpolation.

\section{Strategy for accurate training of PINN}
\label{sct:training}

Previous studies using PINN for travel time in real-world problems, namely GlobeNN \cite[]{Taufik2023SciRep} and HypoSVI \cite[]{Smith2022}, employed a single NN to infer the travel time function in the entire target domain.  
GlobeNN was trained using GLAD-M25 \cite{Lei2020GJI}, a global velocity model derived from global adjoint tomography. The spatial resolution of this model varies with depth but is generally at least 100\,km.
The PINN incorporated into HypoSVI was trained for a relatively small region in Southern California, covering approximately 200\,km$\times$200\,km, horizontally. 
In contrast, the N2018 model covers the entire Nankai Trough subduction zone (approximately 900\,km$\times$300\,km horizontally), derived from marine active-source seismic exploration. 
The typical model resolution beneath the survey lines is on the order of $10^{0}$ to $10^{1}$\,km. 
This scale contrast, which is likely larger than those in the velocity structure models adopted in previous studies, possibly leads to a phenomenon known as ``spectral bias'' \cite[]{Rahaman2019}: 
PINN formulations with fully connected feedforward NNs exhibit poor performance when the target functions include high-frequency or multiscale features \cite[]{Wang2021CMAME}.
Based on preliminary training, we found that the spectral bias can significantly degrade the accuracy of travel time inference near the source ($\leq$ around 100\,km). 
We adopted two approaches to mitigate this effect, namely, domain decomposition and Fourier feature embedding \cite[]{Tancik2020}. 

\begin{figure}
\begin{center}
\begin{small}
\includegraphics[clip, width=8cm, bb = 0 0 424 324]{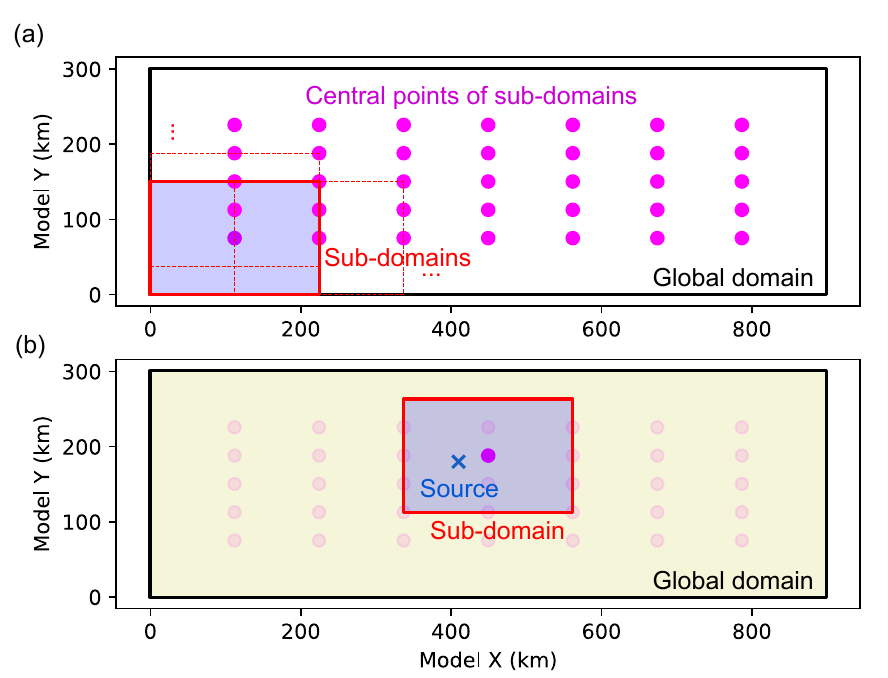}
\end{small}
\end{center}
\caption{Illustration of our simple subdomain-based approach. (a) Subdomain decomposition. The light-blue domain surrounded by the red rectangle is the subdomain of the bottom-left central point. 
(b) Example of source point inference using the subdomain-based approach. The highlighted central point represents the nearest point to the source in the horizontal plane. The light-blue domain surrounded by a red rectangle is the subdomain corresponding to the central point, in which the travel time is inferred using the subdomain NN. The light-yellow domain is the global domain, in which the travel time is inferred by the global domain NN. }
\label{fig:sub-domains}
\end{figure}

The core concept of our domain-decomposition approach is to introduce smaller, additional NNs to more accurately represent the travel time function in a localized area near the source. Limiting the size of the domain make high-frequency components that were present in the original larger domain become mid-frequency components, mitigating the effects of spectral bias \cite[]{Moseley2023ACM}.
The additional NNs were introduced to the travel time function in overlapping subdomains in addition to the NN for the entire domain (hereafter called the global domain). 
Each subdomain spans 225\,km$\times$150\,km in the horizontal direction, with the same thickness as the global domain in the vertical direction. 
We generated five $\times$ seven subdomains in the global domain. 
In inference, the travel time in the subdomain was inferred using the subdomain NN whose horizontal position of the central point was closest to that of the source. 
Those outside the subdomain were inferred using a global-domain NN (Fig. \ref{fig:sub-domains}). 
This is a simpler but similar approach to that of the PINN variations proposed by \cite{Jagtap2020CCP,Moseley2023ACM}, who introduced compatible domain decomposition to the PINN formulation.

The fundamental idea of Fourier feature embedding is to apply a transformation to the input coordinates using a set of sinusoidal functions and encode higher-frequency variations that fully connected feedforward NNs cannot capture efficiently \cite{Tancik2020}.  
We incorporated a trainable version of it \cite[]{Hennigh2021} into each NN (Fig. \ref{fig:VSMNN} (b)).
For NNs in the subdomains, we adopted multiscale Fourier feature embedding \cite[]{Wang2021CMAME} to further improve the convergence (see Supplemental Material).

\section{Training details}
\label{sct:details}

We applied feedforward fully connected NNs with the Swish activation function \cite[]{Ramachandran2018} was used in each layer except for the output function, where linear activation was specified. 
	We introduced NN models of nearly maximum possible size to maintain high expressive power. Studies on image recognition have revealed that, in the absence of additional structures such as skip connections, the maximum number of hidden layers that can be trained without suffering from gradient vanishing is around 20 layers \cite[]{He2016CVPR}. Therefore, we set the number of hidden layers used here to 20. For the global domain, we set the width of the NN to 512, which is nearly the largest size used in previous studies on PINN \cite[]{Wang2023arXiv}. For sub-domains, we set the width to 384, as a smaller size should be sufficient to ensure adequate expressive power. As demonstrated in "Verification through comparison with a grid-based travel time calculation method," these models are efficient enough for rapid hypocenter determination and are suitable for real-time processing despite their large sizes.
The weight parameters of the velocity NN were initialized using He's method \cite[]{He2015}.
The input for the NN were 3D coordinate of the source and receiver locations. 
The trainable parameters in the Fourier feature embeddings were initialized using a zero-mean normal distribution with $\sigma=0.1$. 
The Yogi optimizer \cite[]{Zaheer2018} with an initial learning rate of $3\times10^{-4}$ was used for all the training. 
Furthermore, we used an exponential decay rate of 0.9 every 20,000 steps, following \cite{Wang2023arXiv}. 
All training was conducted with single precision. 

The global-domain NN was trained for 350 epochs with a batch size of 400,000. 
The data size was the same as the number of points in the velocity model after downsampling. 
We used eight NVIDIA A100 GPUs equipped with Earth Simulator, made available by Japan Agency for Marine-Earth Science and Technology (JAMSTEC), for 21 h.
The subdomain NNs were trained for 300 epochs with a batch size of 64,000. 
The data size was the number of points in the velocity model included in the subdomains, which averaged 20 million.
In most cases, two NVIDIA A100 GPUs were used for an average of 12 h to train each NN. Some additional trainings were conducted by using a NVIDIA H100 GPU made available by the TSUBAME4.0 supercomputer at TokyoTech.  
Table \ref{tab:training} summarizes the training details.

\section{Hypocenter determination method}

For hypocenter determination using the P-wave arrival time, we adopted a maximum a posteriori (MAP) estimation, an approach similar to that used by \cite{Hirata1987}. 
This method enables uncertainty quantification from a Bayesian perspective with a Laplace approximation by fitting the posterior probability distribution using a Gaussian distribution. 
The posterior probability density function (PDF) for the hypocenter location ${\bf m}$, namely, longitude, latitude, and depth, is formulated using Bayes' theorem as follows: 
\begin{equation}
P({\bf m}|{\bf d}) \propto P({\bf d} | {\bf m})P({\bf m}), 
\label{eqn:bayestheorem}
\end{equation}
where $P({\bf m}|{\bf d})$, $P({\bf d} | {\bf m})$, and $P({\bf m})$ represent the posterior PDF of the hypocenter parameters, likelihood function, and prior PDF, respectively. 
${\bf d}$ is a data vector, which is the scalar value of the observed arrival time $t^{\rm obs}$ in this specific case. 
We use a likelihood function that eliminates the origin time of the event from the formulation in \cite[]{Ryberg2019}: 
\begin{equation}
P({\bf d} | {\bf m}) = \frac{1}{Z}\exp\left[\sum_{i=1}^{N_{\rm obs}} \frac{\left( \Delta t_{i} - \langle \Delta t \rangle \right)^2}{\sigma_{\rm data}^2} \right],
\end{equation}
where $i$ is an index for observation points, $N_{\rm obs}$ is the number of observation points, $\Delta t_{i}=T_i^{\rm calc}({\bf m})-t_i^{\rm obs}$ and $\langle \Delta t \rangle=\frac{1}{N_{\rm obs}}\sum_{i=1}^{N_{\rm obs}}\Delta t_{i}$. 
$T^{\rm calc}({\bf m})$ is the calculated P-wave travel time dependent on ${\bf m}$, which is rapidly inferred using the PINN-based emulator incorporated in HypoNet Nankai. 
The standard deviation of the data error $\sigma_{\rm data}$ is determined as
\begin{equation}
\sigma_{\rm data}^2 = \sigma_{\rm obs}^2+\sigma_{\rm pred}^2
\end{equation}
where $\sigma_{\rm obs}$ denotes the observation error defined by the user.
$\sigma_{\rm pred}$ represents the model prediction error, which was determined following \cite{Smith2022} by incorporating the error proportional to the travel time observation (see Supplemental Material). 
We use a uniform prior for $P({\bf m})$ with user-specified upper and lower bounds. 

MAP estimation is a nonlinear optimization problem solved using the limited-memory BFGS optimizer \cite[]{Liu1989}. 
During the optimization stage, an invertible logarithmic transform was applied to ${\bf m}$ to incorporate a uniform prior into the gradient-based algorithm \cite[]{Stan2024,Zhang2020Variational}. 
The Hessian matrix for the MAP estimate, which is computable using automatic differentiation available in PyTorch \cite[]{Paszke2017NeurIPS-W}, was used to exploit the approximated analytical expression of the posterior PDF representing estimation uncertainty. 

Additionally, a Python module for PINN-based travel time calculation was made available alongside HypoNet Nankai. Users can utilize this module to implement their own hypocenter determination algorithms.

\section{Verification through comparison with a grid-based travel time calculation method}

\subsection{Verification 1: Travel time prediction}

We set verification problems to compare the travel time inferred by the trained PINN with the results obtained using the fast marching method (FMM) \cite[]{Sethian1996}, which is one of the most widely used grid-based finite-difference numerical calculation methods. Note that Verification 1 targets the trained PINN only and the hypocenter determination algorithm based on the MAP estimation and Laplace approximation are tested in Verification 2.
FMM calculations were conducted using Pykonal \cite[]{White2020SRL}.
We generated random seismic sources within the target 3D domain and computed the travel times for each source at virtual receiver points distributed at 1\,km intervals on the Earth's surface. 
In FMM calculation, the grid intervals were 0.5\,km in the horizontal direction and 0.1\,km in the vertical direction, which were equivalent to the data points in the N2018 velocity model.
These grid intervals have been verified through preliminary runs to be fine enough to provide a reliable reference solution. 

\begin{figure*}
\begin{center}
\begin{small}
\includegraphics[clip, width=16cm, bb = 0 0 890 814]{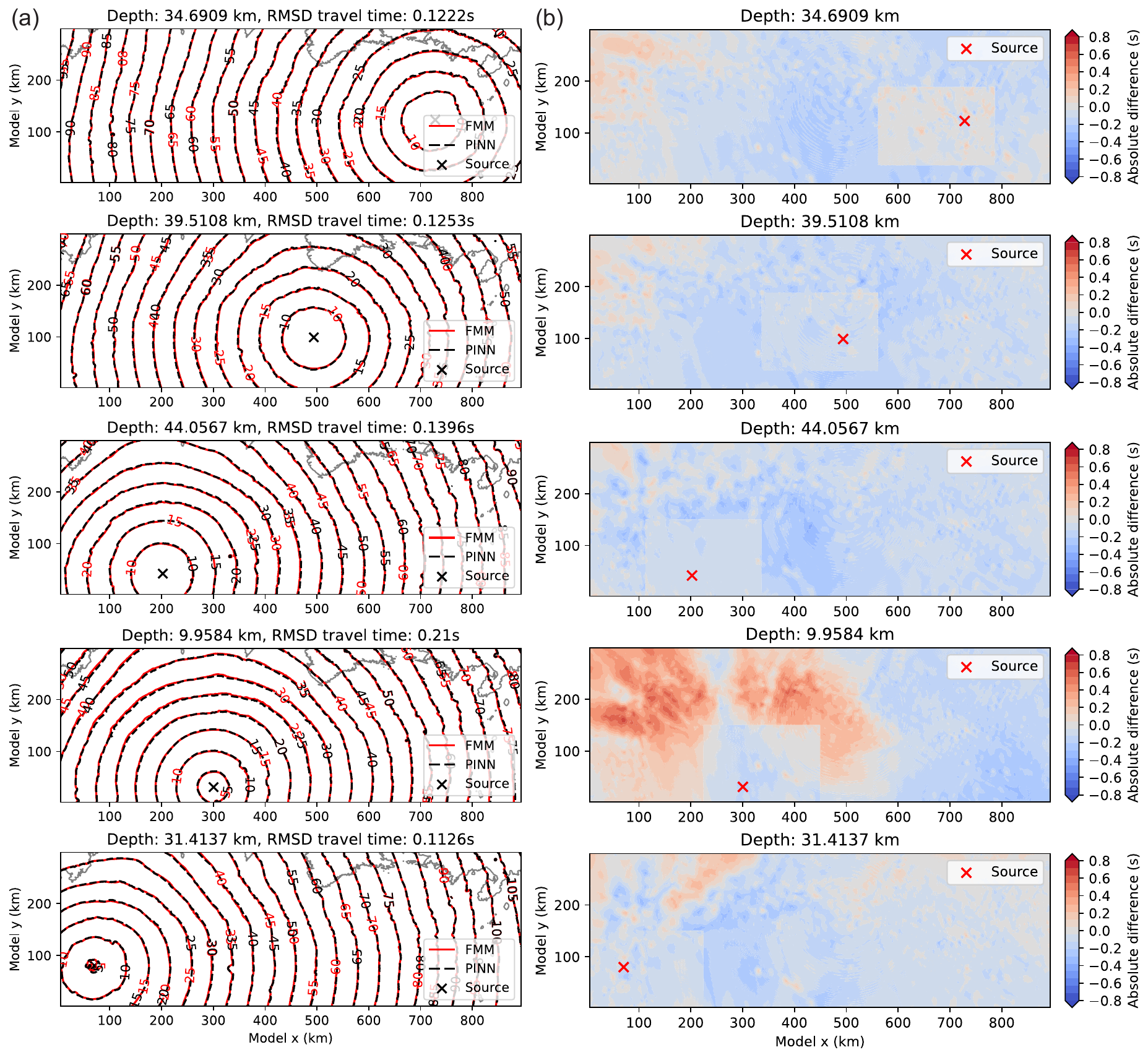}
\end{small}
\end{center}
\caption{Five examples of (a) comparison of calculated (inferred) travel time at the Earth's surface from randomly chosen underground sources obtained using PINN and FMM and (b) their absolute difference (PINN subtracted by FMM). }
\label{fig:tt}
\end{figure*}

\begin{figure*}
\begin{center}
\begin{small}
\includegraphics[clip, width=17cm, bb = 14 0 1078 358]{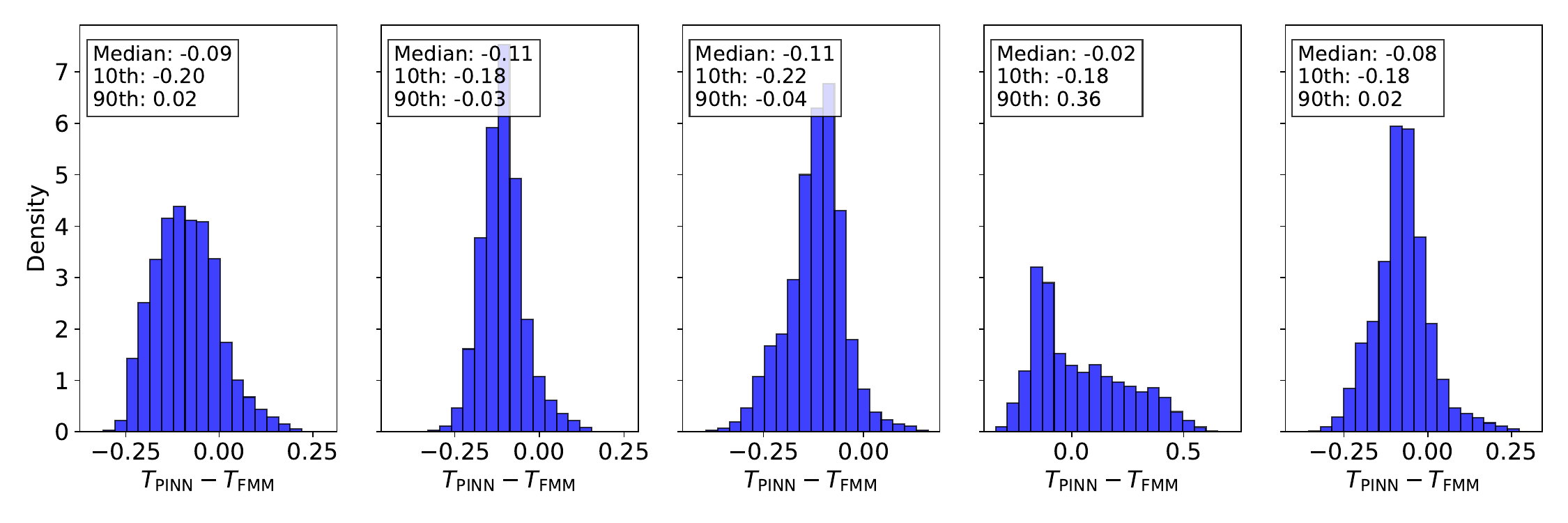}
\end{small}
\end{center}
\caption{Histograms of the point-wisedifference between PINN-inferred and FMM-calculated travel time ($T_{\rm PINN}$ and $T_{\rm FMM}$) for the five source cases presented in Fig. \ref{fig:tt}. Each panel includes statistical information: the median, 10th percentile, and 90th percentile of the distribution.}
\label{fig:hist}
\end{figure*}

The travel time functions on the Earth's surface for five sources, calculated using both PINN and FMM show good agreement at the macroscopic level (Fig. \ref{fig:tt} (a)). 
The root mean square deviation (RMSD) of the travel time in each case is calculated as follows: 
\begin{equation}
\text{RMSD} = \sqrt{\frac{1}{N} \sum_{i=1}^{N} \left( T^{\rm PINN}_{i} - T^{\rm FMM}_{i} \right)^2}, 
\end{equation}
where $T^{\rm PINN/FMM}_{i}$ is the travel time in the $i$-th surface point calculated using either PINN or FMM and $N$ is the number of surface points. The RMSDs are within the range of 0.1 and 0.3\,s. 
Fig. \ref{fig:tt} (a) presents the RMSD for each case. 
A discontinuous decrease in the absolute error in the rectangular region, including the horizontal source location, was apparent for some sources.
This implies that travel time inference using subdomain NNs is more accurate than using the global NN as expected. 
The RMSDs and absolute differences were slightly larger in the fourth case in Fig. \ref{fig:tt} with the shallowest source depth smaller than 10\,km. We plotted the histogram of point-wise differences for all surface points, including the median, 10th percentile, and 90th percentile, for each of the five cases shown in Fig. \ref{fig:hist}. We found a significant bias of the differences toward the negative direction, with a skewed distribution shape, in most cases.  This result indicates that PINN tends to underestimate the travel time slightly, likely due to its tendency to infer smoother travel time function because of spectral bias:
Smoother travel time function results in smaller PINN-inferred travel time gradient, which corresponds to larger predicted velocity $\frac{1}{|\nabla f_{T}(\mathbf{x},\mathbf{x}_{\rm s}; \boldsymbol{\theta})|}$. 
If the predicted velocity is larger than the actual velocity $v(\mathbf{x})$, the inferred travel time is underestimated (see Equation \ref{eqn:loss}). 

\begin{figure}
\begin{center}
\begin{small}
\includegraphics[clip, width=7cm, bb = 14 14 421 350]{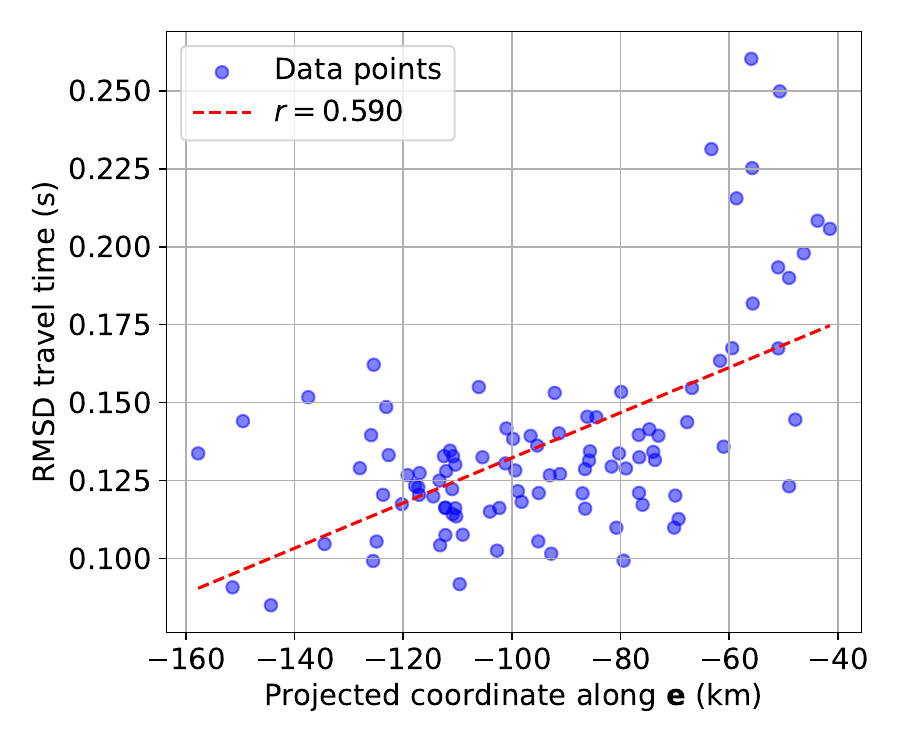}
\end{small}
\end{center}
\caption{The relationship between RMSDs of 100 sources and the projected coordinate along the estimated optimal direction ${\bf e}=(-0.064, -0.190, 0.980)$. The red dashed line indicates the results of linear regression fitted to the data, with the Pearson correlation coefficient $r$ shown in the legend.}
\label{fig:rmsd_projected}
\end{figure}

To gain a more comprehensive understanding of the error characteristics, we repeated the same comparisons for 100 randomly selected sources and calculated the RMSD for each case. Initially, we aimed to determine which of the model axes ($x$, $y$ or $z$) showed the largest correlation with RMSD. However, the calculated correlations for these axes were relatively similar and moderate, the absolute values of which range from 0.3 to 0.4 (Fig. S1). To further refine our analysis, we conducted an optimization to find the spatial axis along which the correlation with RMSD values is maximized. This resulted in a direction vector ${\bf e}=(-0.064, -0.190, 0.980)$ in the model coordinate system and a correlation coefficient of 0.590 (Fig. \ref{fig:rmsd_projected}). 
This result suggests that higher RMSD values tend to correspond to sources located in the shallower ($z$ positive) and seaward ($y$ negative) portions of the velocity model, as exemplified by the fourth source presented in Fig. \ref{fig:tt} (a). These regions are characterized by significant spatial variations in velocity (see Fig. \ref{fig:map} (b) and (c)). This finding indicates that errors in PINN-inferred travel times are generally larger when the source is located in areas with substantial velocity variation.

These comparison results, which show only small differences, imply that our PINN provides results that are as accurate as FMM. 
However, the effect of these differences on the target application of our PINN, that is, hypocenter determination, remains unclear. 
Synthetic experiments to investigate these effects based on hypocenter determination are presented in the following subsection.

\subsection{Verification 2: Numerical experiment of hypocenter determination}
\label{sct:ver2}

We conducted synthetic tests for hypocenter determination using P-wave arrival times, assuming the use of seafloor  seismic observation systems in the Nankai Trough subduction zone. 
We focused on two domains in the Nankai Trough subduction zone where real-time seismological observations are or will be operational.  
Domain 1 is a region with seismometers equipped in Dense Oceanfloor Network system for Earthquakes and Tsunamis (DONET) \cite[]{Kaneda2015,Aoi2020}, and real-time DAS observation points used for the analysis of \cite{Baba2023GRL} off Cape Muroto  (Fig. \ref{fig:map}). 
Domain 2 is the region where Nankai Trough Seafloor Observation Network for Earthquakes and Tsunami (N-net) is being constructed \cite[]{Aoi2023IEEE-UT}. 
We extracted hypocenter models from the 2021 version of the Japanese Meteorological Agency (JMA) earthquake catalog by applying certain magnitude thresholds to ensure a relatively even horizontal distribution of hypocenters.
We calculated the synthetic travel time data between these hypocenter models and the DONET, DAS, and N-net points using FMM, and used them as artificial observation data. 
Artificial noise was not added to the data. 
We then performed hypocenter determination using HypoNet Nankai to determine how the sources were inverted. 
These numerical experiments were conducted separately in Domains 1 and 2. 
Table \ref{tab:domain} summarizes the number of observation points and hypocenter models employed in each experiment. 
Other parameters used in the experiments can be found in Supplemental Material.

\begin{figure*}
\begin{center}
\begin{small}
\includegraphics[clip, width=16cm, bb = 0 0 504 504]{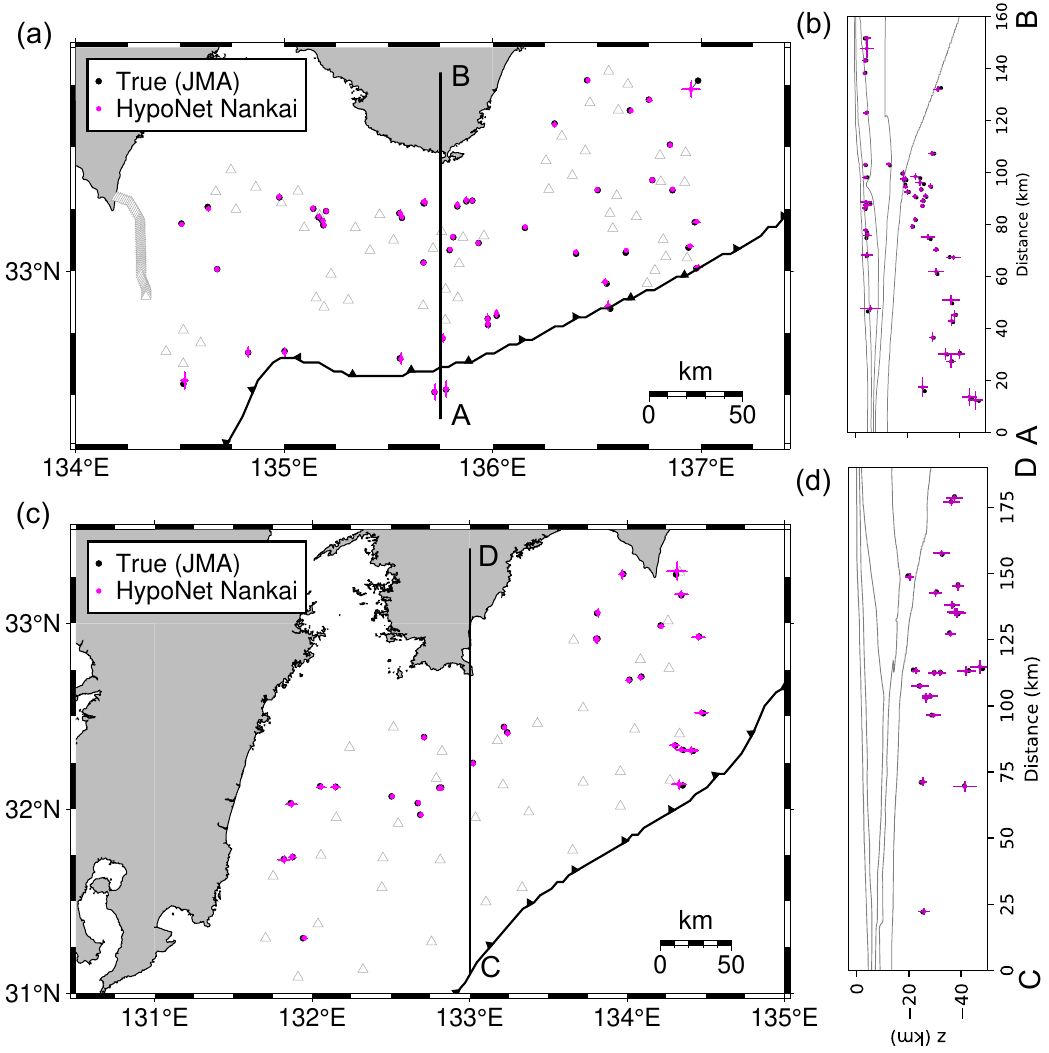}
\end{small}
\end{center}
\caption{Results of the numerical experiments for hypocenter determination using HypoNet Nankai. (a) Results for Domain 1 using DONET and DAS. (b) Plots of cross-section AB described in (a). (c) Results for Domain 2 with N-net. (d) Those plotted on the cross-section CD described in (c). The magenta crossbar indicates the 2-$\sigma$ confidence interval. }
\label{fig:hypo}
\end{figure*}

The MAP-estimated horizontal locations of the hypocenters inside the observation network were accurate and precise in Domains 1 and 2 (Fig. \ref{fig:hypo} (a)(c)). 
The depth components in most points and and the horizontal components in the points outside the observation network tended to show larger estimation uncertainties, which were derived based on the Laplace approximation (Fig. \ref{fig:hypo} (a)(b)(c)(d)). 
Although the differences between the mean  and  true models for some of these events are larger, they fall within the range of estimation uncertainties.
These findings suggest that the inverting models are consistent with the true model locations despite the difference in the forward models used in data generation and inversion, that is, FMM and PINN, respectively. 
The estimation of each hypocenter was completed within two–six seconds using eight CPU cores of AMD EPYC 7742 in Earth Simulator at JAMSTEC. 
Compared to the specifications of a widely  referenced automatic hypocenter determination system \cite[]{Matsumura2006Zisin}, this computation time can be considered sufficiently short to be applicable to such systems.

\section{Discussion}

Once the PINN-based training was completed, HypoNet Nankai required a short computation time for travel time inference in hypocenter determination.
However, an FMM-based approach with an ideal travel-time table created beforehand would require an even shorter time for hypocenter determination when data-loading time for the table is not considered.
Advantages of introducing an NN-based emulator for hypocenter determination lie rather in other aspects.
One advantage of NN-based methods is that they can add new observation points to hypocenter determination systems without much effort. 
For example, among the seafloor seismic observation networks used in this study, the operational time of DONET is a decade longer than that of the DAS system off Muroto.
The operation of N-net has not yet begun.
If we consider to adapt a hypocenter determination system originally created for DONET to DAS and N-net, an FMM-based system would require new travel time calculations to expand the travel-time table. 
In contrast, a system based on HypoNet Nankai can be applied to newly installed points without any modification.  
When planning the installation of new observation points, HypoNet Nankai can be readily applied to feasibility studies of multiple scenario configurations for these new observation points as well.
Another advantage is the size of the dataset. 
The total data size of the 36 NNs included in HypoNet Nankai was approximately 500MB, whereas the FMM table employed in the numerical experiment for Domain 1 alone reached a size of 80\,GB.
This size is not negligible in terms of data-loading time when real-time processing is considered. 
However, reducing the data size sacrifices accuracy.
When developing a tool for public use, developers would never know where potential users would want to place observation points for their calculations. 
Therefore, providing a tool that supports a compact data size and a flexible choice of observation points leveraging NNs is advantageous for this purpose. 

\begin{figure*}
\begin{center}
\begin{small}
\includegraphics[clip, width=16cm, bb = 0 0 504 504]{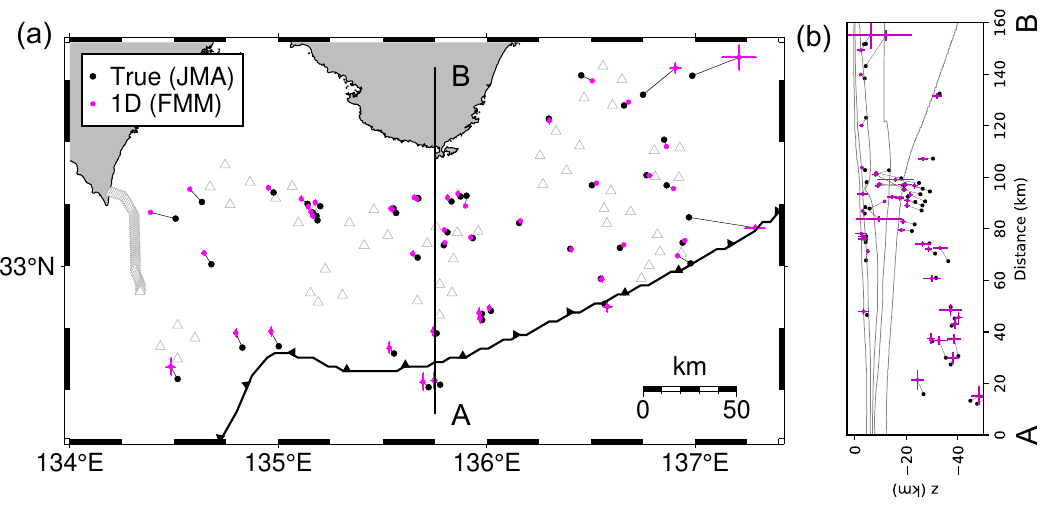}
\end{small}
\end{center}
\caption{Results of numerical experiments of hypocenter determination using 1D velocity structure and FMM. (a) Results for Domain 1 using DONET and DAS. (b) Plots of cross-section AB described in (a).  The magenta crossbar indicates the 2-$\sigma$ confidence interval.}
\label{fig:hypo_1D}
\end{figure*}

To demonstrate the impact of introducing the 3D velocity structure model, we performed a hypocenter determination analysis based on a layered (1D) velocity structure \cite[]{Nakano2013EPS} using the synthetic observational data used in the previous section \ref{sct:ver2} (see Supplemental Material). 
We found a significant discrepancy between the estimated and true hypocenter locations beyond the estimation uncertainty in the horizontal locations of the hypocenters outside the observation network and the depth components of many hypocenters (Fig. \ref{fig:hypo_1D}). 
These patterns of discrepancies are consistent with previous comparisons of hypocenter determinations in 3D versus layered velocity structures \cite[]{Nakano2015,Smith2022}. 
The impact of replacing a simple layered velocity model with a realistic 3D model is evident. 
However, we should also note that the comparison presented here does not reflect the best estimation performance using a 1D model in a practical application. 
This is because station corrections are usually applied to hypocenter determination with 1D models to reduce bias in the estimation originating from model errors. 
From the same perspective, it is important to note that the 3D model introduced in this study is imperfect. 
This includes regions with sparse seismic survey lines, those beyond the reach of the ray path from the active seismic source, and junction points with onshore regions, where the estimation accuracy of the velocity structure is low \cite[]{Nakanishi2018}.
Even in the case of 3D model-based approaches, it may be necessary to implement measures to improve hypocenter determination accuracy, such as estimating station corrections and properly setting model errors in the Bayesian estimation scheme.

\cite{Nakanishi2018} conducted a validation study on the western part of the N2018 P-wave velocity model, which prioritizes marine seismic survey data. Their validation study involved hypocenter determination using real earthquake data and compared the results with another study that determined hypocenters using double-difference tomography directly \cite[]{Yamamoto2013}. However, this validation study suggested that there is still potential for improvement in the model, particularly in the portion beneath the onshore region.
For example, information from models based on the inversion of both seismic survey data and local natural earthquakes with both offshore and onshore first-arrival pick data \cite[]{Arnulf2022NatGeo} and onshore structures that are well validated using waveform simulations \cite[]{Koketsu2009,Koketsu2012} should be incorporated to further improve accuracy. 
These improvements would enable the expansion of the target region for travel time modeling from primarily offshore areas to include onshore regions as well. 
Travel time calculations for the S-wave velocity structure are also essential for better constraining  hypocenters, especially for slow earthquakes \cite[]{Maeda2009}. 
Therefore, constructing an S-wave velocity structure model for the entire domain of the Nankai Trough subduction zone is anticipated. 
Furthermore, HypoNet Nankai should follow these updates to the 3D velocity structure model in the future. 
Notably, if the target region is expanded or if a significant contrast in the seismic velocity are introduced owing to updates in the target velocity structure, the issue of spectral bias may become more pronounced. 
To address this issue, the continuous integration of state-of-the-art techniques to mitigate spectral bias, which image recognition experts are actively developing into PINN (e.g., hash encoding \cite[]{Muller2022ACM,Huang2024JCP}), is essential.

The discontinuity in the travel time function, visible in Fig. \ref{fig:tt} (b), arises at the boundaries where predictions switch between sub-domain and global domain methods. Although this effect was not significant in the current analysis, it could degrade the convergence of the quasi-Newton optimization used by HypoNet Nankai to estimate the hypocenter in some cases.
These discontinuities are inherent to the current domain decomposition approach. To mitigate them and improve accuracy, alternative methods should be considered, such as state-of-the-art techniques for mitigating spectral bias mentioned in the previous paragraph.

\section{Conclusion}

We developed a rapid hypocenter determination tool for the Nankai Trough subduction zone, named HypoNet Nankai, based on a PINN-based emulator for travel time calculation. 
The PINN-based emulator was designed to learn the travel time between arbitrary pairs of sources in a 3D volume and receivers on the Earth's surface in the domain of interest. 
We used a 3D velocity structure model for the Nankai Trough subduction zone that prioritizes marine seismic survey data \cite[]{Nakanishi2018} to train the PINN. 
To cope with the training difficulty due to the small-scale features included within the velocity model, specifically spectral bias, we employed a simple domain decomposition approach and Fourier feature embedding. 
We performed two verification tests to assess the accuracy of the PINN-based emulator and HypoNet Nankai by comparing them against grid-based numerical calculation methods.
In the comparison of forward travel time calculations, we found good agreement in the travel time functions on the Earth's surface for some random sources. 
In the numerical experiment for hypocenter determination, results indicate that hypocenters estimated using HypoNet Nankai are consistent with those obtained using FMM within the range of estimation uncertainty. 
We confirmed that HypoNet Nankai is a fast, accurate, and easy-to-use tool for hypocenter determination for offshore events in the Nankai Trough subduction zones. It is a data-efficient and highly extendible alternative to conventional methods using grid-based travel time calculations. 
HypoNet Nankai is available at \url{http:} (under preparation).

\begin{datres}
Earthquake catalogue used in this study are archived by the Japan Meteorological
Agency (JMA, \url{https://www.data.jma.go.jp/eqev/data/bulletin/hypo_e.html}).
The data usage application for the 3D P-wave velocity structure model of \cite{Nakanishi2018} can be made at \url{https://www.jamstec.go.jp/obsmcs_db/j/Nankai_3D_model-j.html}.
The coordinates of the data points for DAS were made available by \cite{Baba2023GRL} at \url{https://zenodo.org/records/7935235}.
Those for N-net were obtained by digitizing the figure described in \cite{Aoi2023IEEE-UT}.
Supplementary Material provides some technical details on the training of PINN, the hypocenter determination algorithm, and the problem setting of a numerical experiment. 
The software for the review process can be accessed through \url{https://jamstec.box.com/s/afs16i0zhjvdkq0wa35otbvfap6sfc2l}
 (Password: Hyponet1516). 
\end{datres}

\section{Declaration of Competing Interests}

The authors acknowledge that there are no conflicts of interest recorded.

\begin{ack}
This study was supported by JSPS KAKENHI Grant Number 24H01042.
The calculations were carried out using Earth Simulator at JAMSTEC and the TSUBAME4.0 supercomputer at Science Tokyo. 
Some figures were produced using PyGMT \cite[]{Uieda2021}, namely, a Pythonic interface for Generic Mapping Tools 6 (GMT6) \cite[]{Wessel2019}. 
\end{ack}


\begin{thebibliography}{}

\bibitem[\protect\citeauthoryear{Agata, Shiraishi, and Fujie}{Agata et~al.}{2023}]{Agata2023}
Agata, R., K.~Shiraishi, and G.~Fujie (2023).
\newblock {Bayesian Seismic Tomography Based on Velocity-Space Stein Variational Gradient Descent for Physics-Informed Neural Network}.
\newblock {\em IEEE Transactions on Geoscience and Remote Sensing\/}~{\bf 61}, 1--17.

\bibitem[\protect\citeauthoryear{Ando}{Ando}{1975}]{Ando1975}
Ando, M. (1975).
\newblock {Source mechanisms and tectonic significance of historical earthquakes along the Nankai Trough, Japan}.
\newblock {\em Tectonophysics\/}~{\bf 27\/}(2), 119--140.

\bibitem[\protect\citeauthoryear{Aoi, Asano, Kunugi, Kimura, Uehira, Takahashi, Ueda, Shiomi, Matsumoto, and Fujiwara}{Aoi et~al.}{2020}]{Aoi2020}
Aoi, S., Y.~Asano, T.~Kunugi, T.~Kimura, K.~Uehira, N.~Takahashi, H.~Ueda, K.~Shiomi, T.~Matsumoto, and H.~Fujiwara (2020).
\newblock Mowlas: Nied observation network for earthquake, tsunami and volcano.
\newblock {\em Earth, Planets and Space\/}~{\bf 72\/}(1), 1--31.

\bibitem[\protect\citeauthoryear{Aoi, Takeda, Kunugi, Shinohara, Miyoshi, Uehira, Mochizuki, and Takahashi}{Aoi et~al.}{2023}]{Aoi2023IEEE-UT}
Aoi, S., T.~Takeda, T.~Kunugi, M.~Shinohara, T.~Miyoshi, K.~Uehira, M.~Mochizuki, and N.~Takahashi (2023).
\newblock {Development and Construction of Nankai Trough Seafloor Observation Network for Earthquakes and Tsunamis: N-net}.
\newblock In {\em 2023 IEEE Underwater Technology (UT)}, pp.\  1--5. IEEE.

\bibitem[\protect\citeauthoryear{Arnulf, Bassett, Harding, Kodaira, Nakanishi, and Moore}{Arnulf et~al.}{2022}]{Arnulf2022NatGeo}
Arnulf, A.~F., D.~Bassett, A.~J. Harding, S.~Kodaira, A.~Nakanishi, and G.~Moore (2022).
\newblock Upper-plate controls on subduction zone geometry, hydration and earthquake behaviour.
\newblock {\em Nature Geoscience\/}~{\bf 15\/}(2), 143--148.

\bibitem[\protect\citeauthoryear{Baba, Araki, Yamamoto, Hori, Fujie, Nakamura, Yokobiki, and Matsumoto}{Baba et~al.}{2023}]{Baba2023GRL}
Baba, S., E.~Araki, Y.~Yamamoto, T.~Hori, G.~Fujie, Y.~Nakamura, T.~Yokobiki, and H.~Matsumoto (2023).
\newblock {Observation of Shallow Slow Earthquakes by Distributed Acoustic Sensing Using Offshore Fiber-Optic Cable in the Nankai Trough, Southwest Japan}.
\newblock {\em Geophysical Research Letters\/}~{\bf 50\/}(12), e2022GL102678.

\bibitem[\protect\citeauthoryear{{Earthquake Research Committee}}{{Earthquake Research Committee}}{2024}]{ERC2024}
{Earthquake Research Committee} (2024).
\newblock {List of long-term evaluation of earthquake occurrence along major active faults and subduction (in Japanese)}.
\newblock \url{https://www.jishin.go.jp/evaluation/long_term_evaluation/lte_summary}, Accessed 8 July, 2024.

\bibitem[\protect\citeauthoryear{{GEBCO Compilation Group}}{{GEBCO Compilation Group}}{2023}]{Gebco2023}
{GEBCO Compilation Group} (2023).
\newblock {GEBCO 2023 Grid}.

\bibitem[\protect\citeauthoryear{Giroux}{Giroux}{2021}]{Giroux2021}
Giroux, B. (2021).
\newblock ttcrpy: A python package for traveltime computation and raytracing.
\newblock {\em SoftwareX\/}~{\bf 16}, 100834.

\bibitem[\protect\citeauthoryear{Grubas, Duchkov, and Loginov}{Grubas et~al.}{2023}]{Grubas2023}
Grubas, S., A.~Duchkov, and G.~Loginov (2023).
\newblock {Neural Eikonal solver: Improving accuracy of physics-informed neural networks for solving eikonal equation in case of caustics}.
\newblock {\em Journal of Computational Physics\/}~{\bf 474}, 111789.

\bibitem[\protect\citeauthoryear{He, Zhang, Ren, and Sun}{He et~al.}{2015}]{He2015}
He, K., X.~Zhang, S.~Ren, and J.~Sun (2015).
\newblock {Delving deep into rectifiers: Surpassing human-level performance on imagenet classification}.
\newblock In {\em Proceedings of the IEEE international conference on computer vision}, pp.\  1026--1034.

\bibitem[\protect\citeauthoryear{He, Zhang, Ren, and Sun}{He et~al.}{2016}]{He2016CVPR}
He, K., X.~Zhang, S.~Ren, and J.~Sun (2016).
\newblock Deep residual learning for image recognition.
\newblock In {\em 2016 IEEE Conference on Computer Vision and Pattern Recognition (CVPR)}, pp.\  770--778.

\bibitem[\protect\citeauthoryear{Hennigh, Narasimhan, Nabian, Subramaniam, Tangsali, Fang, Rietmann, Byeon, and Choudhry}{Hennigh et~al.}{2021}]{Hennigh2021}
Hennigh, O., S.~Narasimhan, M.~A. Nabian, A.~Subramaniam, K.~Tangsali, Z.~Fang, M.~Rietmann, W.~Byeon, and S.~Choudhry (2021).
\newblock Nvidia simnet™: An ai-accelerated multi-physics simulation framework.
\newblock In {\em International conference on computational science}, pp.\  447--461. Springer.

\bibitem[\protect\citeauthoryear{Hirata and Matsu'ura}{Hirata and Matsu'ura}{1987}]{Hirata1987}
Hirata, N. and M.~Matsu'ura (1987).
\newblock Maximum-likelihood estimation of hypocenter with origin time eliminated using nonlinear inversion technique.
\newblock {\em Physics of the Earth and Planetary Interiors\/}~{\bf 47}, 50--61.

\bibitem[\protect\citeauthoryear{Huang and Alkhalifah}{Huang and Alkhalifah}{2024}]{Huang2024JCP}
Huang, X. and T.~Alkhalifah (2024).
\newblock Efficient physics-informed neural networks using hash encoding.
\newblock {\em Journal of Computational Physics\/}~{\bf 501}, 112760.

\bibitem[\protect\citeauthoryear{Jagtap and Karniadakis}{Jagtap and Karniadakis}{2020}]{Jagtap2020CCP}
Jagtap, A.~D. and G.~E. Karniadakis (2020).
\newblock {Extended physics-informed neural networks (XPINNs): A generalized space-time domain decomposition based deep learning framework for nonlinear partial differential equations}.
\newblock {\em Communications in Computational Physics\/}~{\bf 28\/}(5).

\bibitem[\protect\citeauthoryear{{JAMSTEC}}{{JAMSTEC}}{2004}]{JAMSTEC2004}
{JAMSTEC} (2004).
\newblock {JAMSTEC Seismic Survey Database}.
\newblock \url{https://doi.org/doi:10.17596/0002069}, Accessed 10 June, 2024.

\bibitem[\protect\citeauthoryear{Johnson and Vincent}{Johnson and Vincent}{2002}]{Johnson2002BSSA}
Johnson, M. and C.~Vincent (2002).
\newblock {Development and testing of a 3D velocity model for improved event location: a case study for the India-Pakistan region}.
\newblock {\em Bulletin of the Seismological Society of America\/}~{\bf 92\/}(8), 2893--2910.

\bibitem[\protect\citeauthoryear{Julian, Gubbins, et~al.}{Julian et~al.}{1977}]{Julian1977JG}
Julian, B., D.~Gubbins, et~al. (1977).
\newblock Three-dimensional seismic ray tracing.
\newblock {\em Journal of Geophysics\/}~{\bf 43\/}(1), 95--113.

\bibitem[\protect\citeauthoryear{Kaneda, Kawaguchi, Araki, Matsumoto, Nakamura, Kamiya, Ariyoshi, Hori, Baba, and Takahashi}{Kaneda et~al.}{2015}]{Kaneda2015}
Kaneda, Y., K.~Kawaguchi, E.~Araki, H.~Matsumoto, T.~Nakamura, S.~Kamiya, K.~Ariyoshi, T.~Hori, T.~Baba, and N.~Takahashi (2015).
\newblock {Development and application of an advanced ocean floor network system for megathrust earthquakes and tsunamis}.
\newblock In {\em Seafloor observatories}, pp.\  643--662. Springer.

\bibitem[\protect\citeauthoryear{Katsumata}{Katsumata}{2015}]{Katsumata2015BSSA}
Katsumata, A. (2015).
\newblock {Fast hypocenter determination in an inhomogeneous velocity structure using a 3D travel-time table}.
\newblock {\em Bulletin of the Seismological Society of America\/}~{\bf 105\/}(6), 3203--3208.

\bibitem[\protect\citeauthoryear{Klein}{Klein}{1978}]{Klein1978USGS}
Klein, F.~W. (1978).
\newblock {Hypocenter location program HYPOINVERSE: Part I. Users guide to versions 1, 2, 3, and 4. Part II. Source listings and notes}.
\newblock Technical report, US Geological Survey.

\bibitem[\protect\citeauthoryear{Koketsu, Miyake, and Suzuki}{Koketsu et~al.}{2012}]{Koketsu2012}
Koketsu, K., H.~Miyake, and H.~Suzuki (2012).
\newblock Japan integrated velocity structure model version 1.
\newblock {\em Proceedings of the 15th World Conference on Earthquake Engineering\/}~(1773).

\bibitem[\protect\citeauthoryear{Koketsu, Miyake, Tanaka, et~al.}{Koketsu et~al.}{2009}]{Koketsu2009}
Koketsu, K., H.~Miyake, Y.~Tanaka, et~al. (2009).
\newblock {A proposal for a standard procedure of modeling 3-D velocity structures and its application to the Tokyo metropolitan area, Japan}.
\newblock {\em Tectonophysics\/}~{\bf 472\/}(1-4), 290--300.

\bibitem[\protect\citeauthoryear{Lei, Ruan, Bozda{\u{g}}, Peter, Lefebvre, Komatitsch, Tromp, Hill, Podhorszki, and Pugmire}{Lei et~al.}{2020}]{Lei2020GJI}
Lei, W., Y.~Ruan, E.~Bozda{\u{g}}, D.~Peter, M.~Lefebvre, D.~Komatitsch, J.~Tromp, J.~Hill, N.~Podhorszki, and D.~Pugmire (2020).
\newblock {Global adjoint tomography—model GLAD-M25}.
\newblock {\em Geophysical Journal International\/}~{\bf 223\/}(1), 1--21.

\bibitem[\protect\citeauthoryear{Liu and Nocedal}{Liu and Nocedal}{1989}]{Liu1989}
Liu, D.~C. and J.~Nocedal (1989).
\newblock {On the limited memory BFGS method for large scale optimization}.
\newblock {\em Mathematical programming\/}~{\bf 45\/}(1), 503--528.

\bibitem[\protect\citeauthoryear{Lomax, Zollo, Capuano, and Virieux}{Lomax et~al.}{2001}]{Lomax2001GJI}
Lomax, A., A.~Zollo, P.~Capuano, and J.~Virieux (2001).
\newblock {Precise, absolute earthquake location under Somma--Vesuvius volcano using a new three-dimensional velocity model}.
\newblock {\em Geophysical Journal International\/}~{\bf 146\/}(2), 313--331.

\bibitem[\protect\citeauthoryear{Maeda and Obara}{Maeda and Obara}{2009}]{Maeda2009}
Maeda, T. and K.~Obara (2009).
\newblock {Spatiotemporal distribution of seismic energy radiation from low-frequency tremor in western Shikoku, Japan}.
\newblock {\em Journal of Geophysical Research: Solid Earth\/}~{\bf 114\/}(B10).

\bibitem[\protect\citeauthoryear{Matsumura, Ito, Kimura, Obara, Sekiguchi, Hori, and Kasahara}{Matsumura et~al.}{2006}]{Matsumura2006Zisin}
Matsumura, M., Y.~Ito, H.~Kimura, K.~Obara, S.~Sekiguchi, S.~Hori, and K.~Kasahara (2006).
\newblock {Development of Accurate and Quick Analysis System for Source Parameters (AQUA) (in Japanese with English abstract)}.
\newblock {\em Zisin (Journal of the Seismological Society of Japan. 2nd ser.)\/}~{\bf 59\/}(2), 167--184.

\bibitem[\protect\citeauthoryear{Moseley, Markham, and Nissen-Meyer}{Moseley et~al.}{2023}]{Moseley2023ACM}
Moseley, B., A.~Markham, and T.~Nissen-Meyer (2023).
\newblock Finite basis physics-informed neural networks (fbpinns): a scalable domain decomposition approach for solving differential equations.
\newblock {\em Advances in Computational Mathematics\/}~{\bf 49\/}(4), 62.

\bibitem[\protect\citeauthoryear{Moser}{Moser}{1991}]{Moser1991Geophysics}
Moser, T. (1991).
\newblock Shortest path calculation of seismic rays.
\newblock {\em Geophysics\/}~{\bf 56\/}(1), 59--67.

\bibitem[\protect\citeauthoryear{M\"{u}ller, Evans, Schied, and Keller}{M\"{u}ller et~al.}{2022}]{Muller2022ACM}
M\"{u}ller, T., A.~Evans, C.~Schied, and A.~Keller (2022, jul).
\newblock Instant neural graphics primitives with a multiresolution hash encoding.
\newblock {\em ACM Trans. Graph.\/}~{\bf 41\/}(4).

\bibitem[\protect\citeauthoryear{Nakanishi, Takahashi, Yamamoto, Takahashi, Citak, Nakamura, Obana, Kodaira, and Kaneda}{Nakanishi et~al.}{2018}]{Nakanishi2018}
Nakanishi, A., N.~Takahashi, Y.~Yamamoto, T.~Takahashi, S.~O. Citak, T.~Nakamura, K.~Obana, S.~Kodaira, and Y.~Kaneda (2018).
\newblock {Three-dimensional plate geometry and P-wave velocity models of the subduction zone in SW Japan: Implications for seismogenesis}.
\newblock {\em Geology and Tectonics of Subduction Zones: A Tribute to Gaku Kimura\/}~{\bf 534}, 69.

\bibitem[\protect\citeauthoryear{Nakano, Nakamura, Kamiya, Ohori, and Kaneda}{Nakano et~al.}{2013}]{Nakano2013EPS}
Nakano, M., T.~Nakamura, S.~Kamiya, M.~Ohori, and Y.~Kaneda (2013).
\newblock Intensive seismic activity around the nankai trough revealed by donet ocean-floor seismic observations.
\newblock {\em Earth, planets and space\/}~{\bf 65}, 5--15.

\bibitem[\protect\citeauthoryear{Nakano, Nakamura, and Kaneda}{Nakano et~al.}{2015}]{Nakano2015}
Nakano, M., T.~Nakamura, and Y.~Kaneda (2015).
\newblock {Hypocenters in the Nankai Trough Determined by Using Data from Both Ocean-Bottom and Land Seismic Networks and a 3D Velocity Structure Model: Implications for Seismotectonic Activity}.
\newblock {\em Bulletin of the Seismological Society of America\/}~{\bf 105\/}(3), 1594--1605.

\bibitem[\protect\citeauthoryear{Paszke, Gross, Chintala, Chanan, Yang, DeVito, Lin, Desmaison, Antiga, and Lerer}{Paszke et~al.}{2017}]{Paszke2017NeurIPS-W}
Paszke, A., S.~Gross, S.~Chintala, G.~Chanan, E.~Yang, Z.~DeVito, Z.~Lin, A.~Desmaison, L.~Antiga, and A.~Lerer (2017).
\newblock {Automatic differentiation in PyTorch}.
\newblock In {\em NeurIPS 2017 Workshop Autodiff}.

\bibitem[\protect\citeauthoryear{Pavlis, Holmes, Kenyon, and Factor}{Pavlis et~al.}{2012}]{Pavlis2012}
Pavlis, N.~K., S.~A. Holmes, S.~C. Kenyon, and J.~K. Factor (2012).
\newblock {The development and evaluation of the Earth Gravitational Model 2008 (EGM2008)}.
\newblock {\em Journal of geophysical research: solid earth\/}~{\bf 117\/}(B4).

\bibitem[\protect\citeauthoryear{Rahaman, Baratin, Arpit, Draxler, Lin, Hamprecht, Bengio, and Courville}{Rahaman et~al.}{2019}]{Rahaman2019}
Rahaman, N., A.~Baratin, D.~Arpit, F.~Draxler, M.~Lin, F.~Hamprecht, Y.~Bengio, and A.~Courville (2019).
\newblock On the spectral bias of neural networks.
\newblock In {\em International Conference on Machine Learning}, pp.\  5301--5310. PMLR.

\bibitem[\protect\citeauthoryear{Raissi, Perdikaris, and Karniadakis}{Raissi et~al.}{2019}]{Raissi2019}
Raissi, M., P.~Perdikaris, and G.~E. Karniadakis (2019).
\newblock {Physics-informed neural networks: A deep learning framework for solving forward and inverse problems involving nonlinear partial differential equations}.
\newblock {\em Journal of Computational physics\/}~{\bf 378}, 686--707.

\bibitem[\protect\citeauthoryear{Ramachandran, Zoph, and Le}{Ramachandran et~al.}{2018}]{Ramachandran2018}
Ramachandran, P., B.~Zoph, and Q.~V. Le (2018).
\newblock Searching for activation functions.
\newblock In {\em International Conference on Learning Representations}.

\bibitem[\protect\citeauthoryear{Ryberg and Haberland}{Ryberg and Haberland}{2019}]{Ryberg2019}
Ryberg, T. and C.~Haberland (2019).
\newblock {Bayesian simultaneous inversion for local earthquake hypocentres and 1-D velocity structure using minimum prior knowledge}.
\newblock {\em Geophysical Journal International\/}~{\bf 218\/}(2), 840--854.

\bibitem[\protect\citeauthoryear{Sethian}{Sethian}{1996}]{Sethian1996}
Sethian, J.~A. (1996).
\newblock A fast marching level set method for monotonically advancing fronts.
\newblock {\em proceedings of the National Academy of Sciences\/}~{\bf 93\/}(4), 1591--1595.

\bibitem[\protect\citeauthoryear{Smith, Azizzadenesheli, and Ross}{Smith et~al.}{2021}]{Smith2021}
Smith, J.~D., K.~Azizzadenesheli, and Z.~E. Ross (2021).
\newblock Eikonet: Solving the eikonal equation with deep neural networks.
\newblock {\em IEEE Transactions on Geoscience and Remote Sensing\/}~{\bf 59\/}(12), 10685--10696.

\bibitem[\protect\citeauthoryear{Smith, Ross, Azizzadenesheli, and Muir}{Smith et~al.}{2022}]{Smith2022}
Smith, J.~D., Z.~E. Ross, K.~Azizzadenesheli, and J.~B. Muir (2022).
\newblock {HypoSVI: Hypocentre inversion with Stein variational inference and physics informed neural networks}.
\newblock {\em Geophysical Journal International\/}~{\bf 228\/}(1), 698--710.

\bibitem[\protect\citeauthoryear{{Stan Development Team}}{{Stan Development Team}}{2024}]{Stan2024}
{Stan Development Team} (2024).
\newblock {Stan Modeling Language Users Guide and Reference Manual, Version 2.35}.
\newblock \url{https://mc-stan.org}, Accessed 16 July, 2024.

\bibitem[\protect\citeauthoryear{Tancik, Srinivasan, Mildenhall, Fridovich-Keil, Raghavan, Singhal, Ramamoorthi, Barron, and Ng}{Tancik et~al.}{2020}]{Tancik2020}
Tancik, M., P.~Srinivasan, B.~Mildenhall, S.~Fridovich-Keil, N.~Raghavan, U.~Singhal, R.~Ramamoorthi, J.~Barron, and R.~Ng (2020).
\newblock {Fourier features let networks learn high frequency functions in low dimensional domains}.
\newblock {\em Advances in Neural Information Processing Systems\/}~{\bf 33}, 7537--7547.

\bibitem[\protect\citeauthoryear{Taufik, Waheed, and Alkhalifah}{Taufik et~al.}{2023}]{Taufik2023SciRep}
Taufik, M.~H., U.~b. Waheed, and T.~A. Alkhalifah (2023).
\newblock {A neural network based global traveltime function (GlobeNN)}.
\newblock {\em Scientific Reports\/}~{\bf 13\/}(1), 7179.

\bibitem[\protect\citeauthoryear{Uieda, Tian, Leong, Toney, Schlitzer, Yao, Grund, Jones, Materna, Newton, Ziebarth, and Wessel}{Uieda et~al.}{2021}]{Uieda2021}
Uieda, L., D.~Tian, W.~J. Leong, L.~Toney, W.~Schlitzer, J.~Yao, M.~Grund, M.~Jones, K.~Materna, T.~Newton, M.~Ziebarth, and P.~Wessel (2021, March).
\newblock {PyGMT: A Python interface for the Generic Mapping Tools}.

\bibitem[\protect\citeauthoryear{Um and Thurber}{Um and Thurber}{1987}]{Um1987BSSA}
Um, J. and C.~Thurber (1987).
\newblock A fast algorithm for two-point seismic ray tracing.
\newblock {\em Bulletin of the Seismological Society of America\/}~{\bf 77\/}(3), 972--986.

\bibitem[\protect\citeauthoryear{Waheed, Haghighat, Alkhalifah, Song, and Hao}{Waheed et~al.}{2021}]{Waheed2021PINNeik}
Waheed, U.~B., E.~Haghighat, T.~Alkhalifah, C.~Song, and Q.~Hao (2021).
\newblock {PINNeik: Eikonal solution using physics-informed neural networks}.
\newblock {\em Computers \& Geosciences\/}~{\bf 155}, 104833.

\bibitem[\protect\citeauthoryear{Wang, Sankaran, Wang, and Perdikaris}{Wang et~al.}{2023}]{Wang2023arXiv}
Wang, S., S.~Sankaran, H.~Wang, and P.~Perdikaris (2023).
\newblock {An expert's guide to training physics-informed neural networks}.
\newblock {\em arXiv preprint arXiv:2308.08468\/}.

\bibitem[\protect\citeauthoryear{Wang, Wang, and Perdikaris}{Wang et~al.}{2021}]{Wang2021CMAME}
Wang, S., H.~Wang, and P.~Perdikaris (2021).
\newblock {On the eigenvector bias of Fourier feature networks: From regression to solving multi-scale PDEs with physics-informed neural networks}.
\newblock {\em Computer Methods in Applied Mechanics and Engineering\/}~{\bf 384}, 113938.

\bibitem[\protect\citeauthoryear{Wessel, Luis, Uieda, Scharroo, Wobbe, Smith, and Tian}{Wessel et~al.}{2019}]{Wessel2019}
Wessel, P., J.~Luis, L.~Uieda, R.~Scharroo, F.~Wobbe, W.~H. Smith, and D.~Tian (2019).
\newblock The generic mapping tools version 6.
\newblock {\em Geochemistry, Geophysics, Geosystems\/}~{\bf 20\/}(11), 5556--5564.

\bibitem[\protect\citeauthoryear{White, Fang, Nakata, and Ben-Zion}{White et~al.}{2020}]{White2020SRL}
White, M.~C., H.~Fang, N.~Nakata, and Y.~Ben-Zion (2020).
\newblock {PyKonal: a Python package for solving the eikonal equation in spherical and Cartesian coordinates using the fast marching method}.
\newblock {\em Seismological Research Letters\/}~{\bf 91\/}(4), 2378--2389.

\bibitem[\protect\citeauthoryear{Yamamoto, Obana, Takahashi, Nakanishi, Kodaira, and Kaneda}{Yamamoto et~al.}{2013}]{Yamamoto2013}
Yamamoto, Y., K.~Obana, T.~Takahashi, A.~Nakanishi, S.~Kodaira, and Y.~Kaneda (2013).
\newblock {Imaging of the subducted Kyushu-Palau Ridge in the Hyuga-nada region, western Nankai Trough subduction zone}.
\newblock {\em Tectonophysics\/}~{\bf 589}, 90--102.

\bibitem[\protect\citeauthoryear{Yamamoto, Obana, Takahashi, Nakanishi, Kodaira, and Kaneda}{Yamamoto et~al.}{2014}]{Yamamoto2014}
Yamamoto, Y., K.~Obana, T.~Takahashi, A.~Nakanishi, S.~Kodaira, and Y.~Kaneda (2014).
\newblock {Seismicity and structural heterogeneities around the western Nankai Trough subduction zone, southwestern Japan}.
\newblock {\em Earth and Planetary Science Letters\/}~{\bf 396}, 34--45.

\bibitem[\protect\citeauthoryear{Yamamoto, Takahashi, Kaiho, Obana, Nakanishi, Kodaira, and Kaneda}{Yamamoto et~al.}{2017}]{Yamamoto2017EPSL}
Yamamoto, Y., T.~Takahashi, Y.~Kaiho, K.~Obana, A.~Nakanishi, S.~Kodaira, and Y.~Kaneda (2017).
\newblock {Seismic structure off the Kii Peninsula, Japan, deduced from passive-and active-source seismographic data}.
\newblock {\em Earth and Planetary Science Letters\/}~{\bf 461}, 163--175.

\bibitem[\protect\citeauthoryear{Zaheer, Reddi, Sachan, Kale, and Kumar}{Zaheer et~al.}{2018}]{Zaheer2018}
Zaheer, M., S.~Reddi, D.~Sachan, S.~Kale, and S.~Kumar (2018).
\newblock Adaptive methods for nonconvex optimization.
\newblock In S.~Bengio, H.~Wallach, H.~Larochelle, K.~Grauman, N.~Cesa-Bianchi, and R.~Garnett (Eds.), {\em Advances in Neural Information Processing Systems}, Volume~31. Curran Associates, Inc.

\bibitem[\protect\citeauthoryear{Zhang and Curtis}{Zhang and Curtis}{2020}]{Zhang2020Variational}
Zhang, X. and A.~Curtis (2020).
\newblock Variational full-waveform inversion.
\newblock {\em Geophysical Journal International\/}~{\bf 222\/}(1), 406--411.

\bibitem[\protect\citeauthoryear{Zhao}{Zhao}{2005}]{Zhao2005}
Zhao, H. (2005).
\newblock A fast sweeping method for eikonal equations.
\newblock {\em Mathematics of computation\/}~{\bf 74\/}(250), 603--627.

\end{thebibliography}

\vspace{2cm}
The mailing address of all the authors is: 
\begin{itemize}
    \item 3173-25, Showa-machi, Kanazawa-ku, Yokohama-city, Kanagawa, 236-0001, Japan
\end{itemize}

\clearpage

\begin{table}[!b]
\tbl{Summary of the hyperparameters and computational resources employed in the training of the PINN-based emulator for HypoNet Nankai.\label{tab:training}}
{\begin{tabular}{|l|l|l|} \hline %
   Parameters & Sub-domains & Global domain \\ \hline\hline
   \# hidden layers & 20 & 20 \\ \hline
   \# neurons per layer & 384 & 512 \\ \hline
   Encoding & Multi-scale FF & Trainable FF \\\hline
   Activation function & \multicolumn{2}{c|}{Swish} \\\hline
   Optimization algorithm & \multicolumn{2}{c|}{Yogi} \\ \hline
   Initial learning rate & \multicolumn{2}{c|}{$3\times10^{-4}$} \\ \hline
   \# Data points & $\sim$20,000,000 & 156,196,312  \\ \hline
   Batch size & 64,000 & 400,000 \\ \hline
   \# epochs & 300 & 350 \\ \hline
   Hardware  & A100 $\times$ 2 & A100 $\times$ 8 \\ \hline
   Training time & $\sim$12\,hrs  & 21\,hrs \\ \hline
\end{tabular}}
\end{table}

\begin{table}
\tbl{Summary of the number of hypocenters and observation points employed in the numerical experiments for hypocenter determination. .\label{tab:domain}}
{\begin{tabular}{|l|c|c|} \hline %
    & Domain 1 & Domain 2 \\ \hline\hline
   \# hypocenters & 47 & 30 \\ \hline
   \# observation points & 106 & 36 \\ \hline
\end{tabular}}
\end{table}

\end{document}